\documentclass[10pt,letterpaper]{article}
\usepackage{opex3}
\usepackage{graphicx}

\begin{document}

%%%%%%%%%%%%%%%%%% title page information %%%%%%%%%%%%%%%%%%
\title{Probing higher order correlations of the photon field with photon number resolving avalanche photodiodes}

\author{J. F. Dynes, Z. L. Yuan, A. W. Sharpe, O. Thomas and A. J. Shields}

\address{Toshiba Research Europe Limited, Cambridge Research Laboratory, 208 Cambridge Science Park, Milton Road, Cambridge, CB4 0GZ, UK}

\email{james.dynes@crl.toshiba.co.uk} %% email address is required

% \homepage{http:...} %% author's URL, if desired

%%%%%%%%%%%%%%%%%%% abstract and OCIS codes %%%%%%%%%%%%%%%%
%% [use \begin{abstract*}...\end{abstract*} if exempt from copyright]

\begin{abstract} We demonstrate the use of two high speed avalanche photodiodes in exploring higher order photon correlations. By employing the photon number resolving capability of the photodiodes the response to higher order photon coincidences can be measured. As an example we show experimentally the sensitivity to higher order correlations for three types of photon sources with distinct photon statistics. This higher order correlation technique could be used as a low cost and compact tool for quantifying the degree of correlation of photon sources employed in quantum information science.\end{abstract}

 % REPLACE WITH CORRECT OCIS CODES FOR YOUR ARTICLE

%%%%%%%%%%%%%%%%%%%%%%% References %%%%%%%%%%%%%%%%%%%%%%%%%

%%%%%%%%%%%%%%%%%%%%%%%%%%  body  %%%%%%%%%%%%%%%%%%%%%%%%%%
\section{Introduction}
Photon correlation functions, formalized by Glauber in the 1960s \cite{glauber1963}, have become fundamental tools in  characterizing the quantum statistical properties of light. From an experimental point of view, it is the class of normally ordered correlation functions \cite{milburn2008} corresponding to photon absorption as opposed to the anti-normally ordered class \cite{usami2004} that feature as most accessible. Up until now, the most significant normally ordered photon correlation function is the second order correlation function, first measured by Hanbury-Brown and Twiss (HBT) \cite{hanbury1956,jeltes2007}. For example, the second order correlation function makes possible photon states to be categorized as classical or quantum mechanical. Poissonian or bunched photon statistics would indicate classical behavior - or if sub-Poissonian statistics are observed then it would be concluded the photon field had a quantum character \cite{kimble1977}.

Despite the strong interest in measuring the second order correlation function, comparatively little research to date has focused on measuring higher order correlations. Partly this is due to the complexity of the task at hand; characterizing higher order correlations of a given light source can require as many detectors as the order of the correlation function being measured itself \cite{agarwal1970}. Nevertheless, from both a practical and fundamental point of view, characterization of higher order correlation functions is of considerable importance. In view of applications, for example quantum key distribution (QKD) \cite{gisin2002_gisin2007}, statistical knowledge of the source is vital for security. Furthermore, it has been shown for QKD using realistic single photon sources such as semiconductor quantum dots \cite{michler2000,santori2001,yuan2002}, the secure transmission distance can approach that of using a theoretical, true single photon source with knowledge of the source higher order photon correlation functions \cite{adachi2009}.

Traditionally, monitoring the photon field statistics by measuring higher order correlations is a time consuming process \cite{agarwal1970}. The time taken to acquire a large number of coincidences to achieve acceptable statistical bounds on the correlation function scales exponentially with the order of the correlation. This is usually impractical in many situations. Alternative techniques for measuring higher order correlations have been put forward such as two-photon detection \cite{qu1996}, spatial multiplexing of superconducting nanowire detectors \cite{stevens2009} and temporal multiplexing based on linear optics \cite{avenhaus2009}. While these techniques have their own individual merits, they suffer from other significant drawbacks including electrical cross-talk, cryogenic operating temperatures and low coincidence rates. Moreover, to measure higher order correlations where the order is greater than about four, the above techniques may prove problematic due to scalability issues. We note that very recently a proposal to filter out cross-talk in multi-pixel detectors for accessing higher order correlations has been reported \cite{kalashnikov2011}.

Measuring higher order correlations with just two detectors is a relatively unexplored area. One possible method is to make use of the detector photon number resolving capability. Most single photon detectors though can only differentiate between the zero photons and one or more photons \cite{hadfield2009}.  A single or multi-photon event produces the same voltage pulse height, making photon number resolving impossible. On the other hand, photon number resolving (PNR) detectors produce voltage pulses in proportion to the number of photons. Hence, $n$-photon events can be precisely selected depending on the discrimination levels. In this paper, we present a proof of principle experiment for demonstrating sensitivity to higher order correlations based on high speed PNR detectors. We employ two gigahertz gated, PNR avalanche photodiodes (APDs) in a normal HBT setup \cite{hanbury1956,jeltes2007}. This setup can measure the usual normally ordered and normalized second order correlation function $g$(2) defined as $g(2) = \frac{<\hat{a}^{\dagger 2}\hat{a}^{2}>}{<\hat{a}^{\dagger}\hat{a}>^2}$ where $\hat{a}^{\dagger}$ and $\hat{a}$ are the usual photon creation and annihilation operators. The form of $g(2)$ presented above is the most salient in terms of categorizing photon states. However, we demonstrate that due to the PNR capability of the detectors, two-photon, three photon and up to $n$-photon events can be selected using photon counting discriminators, where $n$ is the PNR capability of the detector. Coupled with the joint photon detection of using both PNR detectors together, the setup is sensitive to higher order correlations. As we show below for a bunched photon source, the sensitivity to photon bunching is significantly improved over the standard $g(2)$ measurement.

\section{Theory}

According to Glauber \cite{glauber1963}, the definition of the $n$th order, normally ordered, normalized correlation function is given by:
\begin{equation}\label{eq:glauber}
g(n) = \frac{<\hat{a}^{\dagger n}\hat{a}^{n}>}{<\hat{a}^{\dagger}\hat{a}>^n}
\end{equation}
When $n=2$, $g(2)$ can be measured using the usual Hanbury-Brown Twiss experiment with a double single photon detector arrangement. Such a setup is depicted in Fig. \ref{fig:fig1.eps}(a) with two InGaAs APDs as the detectors. In this case, both detector discriminators are set so 0-photons are rejected while 1-photon and multi-photon events are sampled. This is readily achieved using most non-photon number resolving single photon detectors. 
As an example, if the source under investigation were purely thermal with a coherence time greater than the temporal resolution of the detector, a standard HBT experiment would yield $g(2) = 2$. Specifically, the measurement involves sampling the coincident counts at times smaller than the coherence time of the source and dividing by the accidental coincidences (obtained at times greater than the source coherence time).

If the detectors are photon number resolving, then information on higher order correlation functions can be obtained by adjusting the discrimination levels. Sampling only 2-photon events by rejecting 0, 1 and $\geq$ 3 photon events can be achieved with window discriminators with the window set to count only 2-photon events.
We can define a quantity called the higher order coincidence, for example $\gamma(4)$:
\begin{equation}\label{eq:excess}
\gamma(4) = \frac{<\hat{a}^{\dagger4}\hat{a}^4>}{<\hat{a}^{\dagger 2}\hat{a}^2><\hat{a}^{\dagger 2}\hat{a}^2>}
\end{equation}
$\gamma(4)$ clearly contains information regarding higher order correlation functions as we can re-write $\gamma(4)$ in terms of the usual normalized correlation functions:
\begin{equation}\label{eq:gamma_4}
\gamma(4) = \frac{g(4)}{g(2)*g(2)}
\end{equation} 
In the example given above with the thermal source, the higher order coincidence rate would be higher than $g(2)$ with $\gamma(4)=6$.

The above analysis can be extended to higher photon numbers. We can generalize Eq. (\ref{eq:excess}) as:
\begin{equation}\label{eq:eq1}
\gamma(n_{1}+n_{2}) = \frac{<\hat{a}^{\dagger(n_{1}+n_{2})}\hat{a}^{(n_{1}+n_{2})}>}{<\hat{a}^{\dagger n_{2}}\hat{a}^{n_{2}}><\hat{a}^{\dagger n_{1}}\hat{a}^{n_{1}}>}
\end{equation}
Here $n_{i}$ indicates that the window discrimination levels of detector $i$ is placed around the $n_{i}$ photon number states. Eq. (\ref{eq:eq1}) can also be re-written in terms of normalized correlation functions:
\begin{equation}\label{eq:gamma_n1_n2}
\gamma(n_{1}+n_{2}) = \frac{g(n_{1}+n_{2})}{g(n_{1})*g(n_{2})}
\end{equation} 
If the photon number resolution of both detectors is $n$, then the higher order coincidence can be measured up to $2n$.  Note that this arrangement circumvents the non-unity detector efficiency of the detectors and any inherent optical losses in the system.

For simplicity and to illustrate the proof of principle experiment presented below, we use threshold discriminators rather than window discriminators.  To incorporate that fact it is necessary to integrate over all photon number states greater than the lowest allowed photon number state. In this case Eq. (\ref{eq:eq1}) is modified to read:
\begin{equation}\label{eq:eq2}
\gamma(n_{1}+n_{2}) = \frac{\sum^{n_{max}}_{n_1,n_2}<\hat{a}^{\dagger(n_{1}+n_{2})}\hat{a}^{(n_{1}+n_{2})}>\eta^{n_1}_{1}\eta^{n_2}_{2}}{\sum^{n_{max}}_{n_1,n_2}<\hat{a}^{\dagger n_{2}}\hat{a}^{n_{2}}><\hat{a}^{\dagger n_{1}}\hat{a}^{n_{1}}>\eta^{n_1}_{1}\eta^{n_2}_{2}}
\end{equation}
where $n_{max}$ is the maximum photon number the detector can measure and $\eta_{i}$ is the $i$ system single photon detection efficiency. The photon sources appropriate for use with this technique have detected mean photon fluxes much less than $2n_{max}$. Note that Eq. (\ref{eq:eq2}) approximates Eq. (\ref{eq:eq1}) when $\eta_{i}$ is relatively small. This is due to the weighting of the higher order terms in the sums of Eq. (\ref{eq:eq2}) scaling as $\eta^{n_i}_{i}$. 

\section{Photon number resolution with InGaAs avalanche photodiodes}
\begin{figure}[b!]
{\centerline{\scalebox{0.45}{\includegraphics{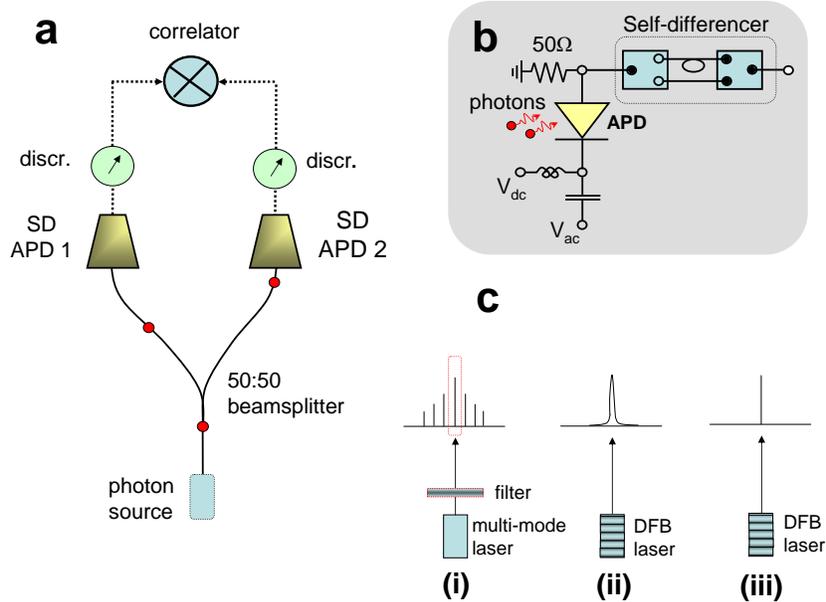}}}}
\caption{\label{fig:fig1.eps}(a) Schematic of the experimental setup. The solid circles represent photons from the source under test. SD-APD1 \& SD-APD2: self-differencing avalanche photodiodes (APDs), discr.: electrical signal discriminator.{(b)} Electrical setup for measuring weak avalanches. {(c)} The three types of photon source employed and pictorial representations of their frequency spectra: (i) Filtered multi-mode laser operating slightly above lasing threshold (FML) (the mode in red box denotes allowed mode), (ii) DFB laser operating near lasing threshold (LNT) \& (iii) DFB laser operated well above threshold (LAT).}
\end{figure}

\begin{figure}[t!]
{\centerline{\scalebox{0.45}{\includegraphics{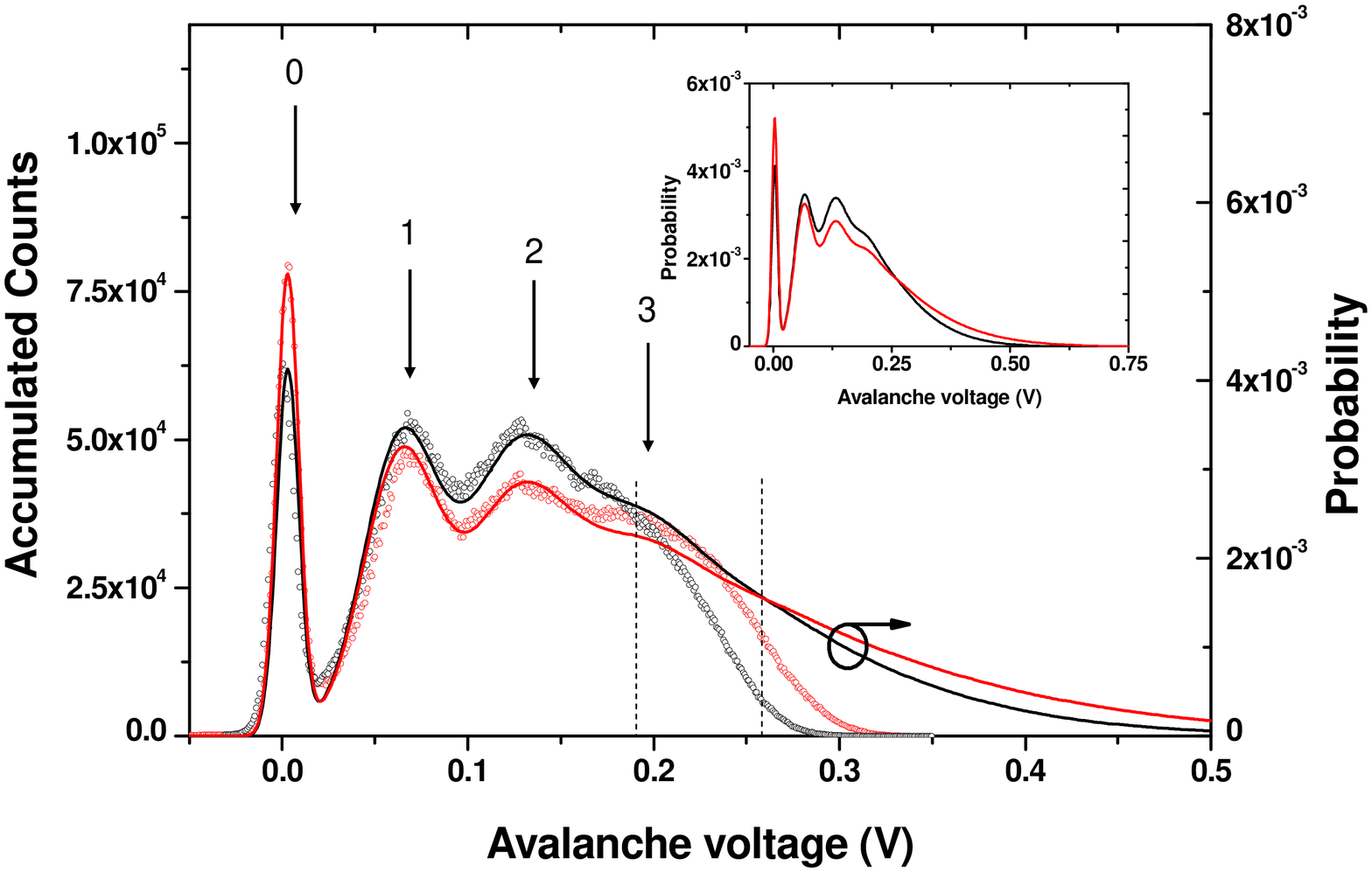}}}}
\caption{\label{fig:fig2.eps} The experimentally measured distribution of avalanche voltages for a pulsed laser source operating well above threshold LAT (black circles) and for a pulsed, mode filtered non-DFB laser source operating close to threshold FML (red circles). Also shown are the theoretical distributions of avalanches for a Poissonian source (black line) and a partially bunched source (red line).  The dashed lines correspond to the cross-over avalanche voltages for the experimental and theoretical avalanche distributions, as described in the text. Inset: Depicts the theoretical avalanche distributions for the Poissonian source (black line) and the partially bunched source (red line) up to an avalanche voltage of 0.75V.}
\end{figure}

Fig. \ref{fig:fig1.eps}(a) shows the layout of the experimental setup. The photon source under test is coupled into single mode fiber and photons propagate along the fiber to a 50:50 fiber beamsplitter. Photons in each arm are detected by an InGaAs APD operating in a self-differencing mode at a clock rate of 1GHz \cite{yuan2007}. The self-differencer permits cancelation of the strong APD capacitive response leaving behind the previously obscured weak avalanche. This is achieved by dividing the electrical APD output into two arms, delaying one arm by an integer number of clock cycles before recombining the two arms using a subtractive combiner, Fig. \ref{fig:fig1.eps}(b). After self-differencing, the avalanches from the detectors are amplified before being discriminated by ultra-fast (1ns) discriminators. The discriminated outputs are then sent to a two channel time-correlated photon counting card (correlator) for recording the photon arrival times. For measuring the detector avalanche voltage distribution, a high speed oscilloscope is employed directly on the amplified detector output. The APDs feature single photon detection efficiencies of around 15--20\%.

Fig. \ref{fig:fig2.eps} shows the experimentally measured avalanche voltage distribution of a detector for two types of photon source. When excited with a $1550$nm distributed feedback (DFB) type pulsed laser operating well above threshold  (LAT, Fig. \ref{fig:fig1.eps}(c)), the avalanche voltage distribution shows a number of distinct peaks, black circles Fig. \ref{fig:fig2.eps}. These peaks correspond to different photon number, $n$ in the incident light field. The avalanche voltage distribution can be satisfactorily modeled assuming the photon number distribution of the source is Poissonian with a mean detected photon flux, $\mu \sim 2.6$, black line in Fig. \ref{fig:fig2.eps}. Each photon peak, $n\geq1$ is assumed to be Gaussian and to reflect the statistical broadening due to avalanche noise. The widths of the photon peaks are scaled as $\sqrt{n}$ relative to the 1-photon peak width. Such a technique has successfully been employed to model photon number distributions for both InGaAs and Silicon APDs \cite{kardynal2008,thomas2010}. The areas of each photon peak are proportional to the expected Poisson photon number distribution probabilities \cite{kardynal2008}. The main features of the LAT trace (black circles) in Fig. \ref{fig:fig2.eps} are reproduced by the black line.

The ability to detect differences in the photon statistics for dissimilar photon sources is now illustrated.
The photon source was replaced with a multi-mode pulsed laser operating slightly above threshold with a $1$nm bandpass filter that serves to pass a single spectral mode close to a wavelength of $1550$nm (FML in Fig. \ref{fig:fig1.eps}(c)(i)). Here the average detected photon flux was similar to before with $\mu \sim 2.8$.

This photon source gives rise to a rather different avalanche voltage distribution, (red circles, Fig. \ref{fig:fig2.eps}). The 0-photon peak is higher than for the laser source operating well above threshold, indicating the presence of more vacuum states. Conversely there is a suppression of the 2-photon and 3-photon number states. The photon source is expected to exhibit intensity fluctuations due to operation near to lasing threshold. Such fluctuation is manifest experimentally as photon bunching \cite{mandel1995}. Bunching can be viewed as a departure from the independent statistics of a Poissonian source. The degree of bunching is accentuated through selection of a single mode by filtering \cite{dixon2009}. The mathematical form of the photon number distribution for this particular case is not known precisely, so we choose an analytic photon counting distribution function representing a linear superposition of a coherent state field with mean photon number $\mu_{s}$ and a chaotic field with mean photon number $\mu_n$\cite{smith1966,lachs1965}. 
The following relations fix $\mu_s$ \& $\mu_n$ from experimentally measurable quantities, namely average detected photon flux, $\mu =\mu_{s} + \mu_{n}$ and $g(2) = \mu_{n}(\mu_{n} + 2\mu_{s})/\mu^2 +1$ \cite{smith1966,lachs1965}. Using $\mu=2.8$ and $g(2)=1.2$ \cite{g2ref}, the resulting modeled avalanche distribution is shown in Fig. \ref{fig:fig2.eps} (red line). 

For both types of photon source, the theoretical avalanche distribution overestimates large avalanche heights. 
The origin of this effect is unclear at present. However, it is quite likely due to the quenching by the APD series resistance causing large avalanches from photon number states with $n > 3$ to be reduced in size (the linear model assumes a linear avalanche voltage dependence on photon number). The quenching effect can be somewhat quantified by comparing the experimental curves crossover point with the theoretical crossover point. These occur at 0.18 and 0.26V respectively, as indicated by the dashed lines in Fig. \ref{fig:fig2.eps}. Hence there is a 30\% reduction in avalanche height at an avalanche voltage of 0.26V compared to the simple linear model prediction. %We emphasize the above discrepancies do not play a role in the correlation measurements described below. 

\begin{figure}[t!]
{\centerline{\scalebox{0.6}{\includegraphics{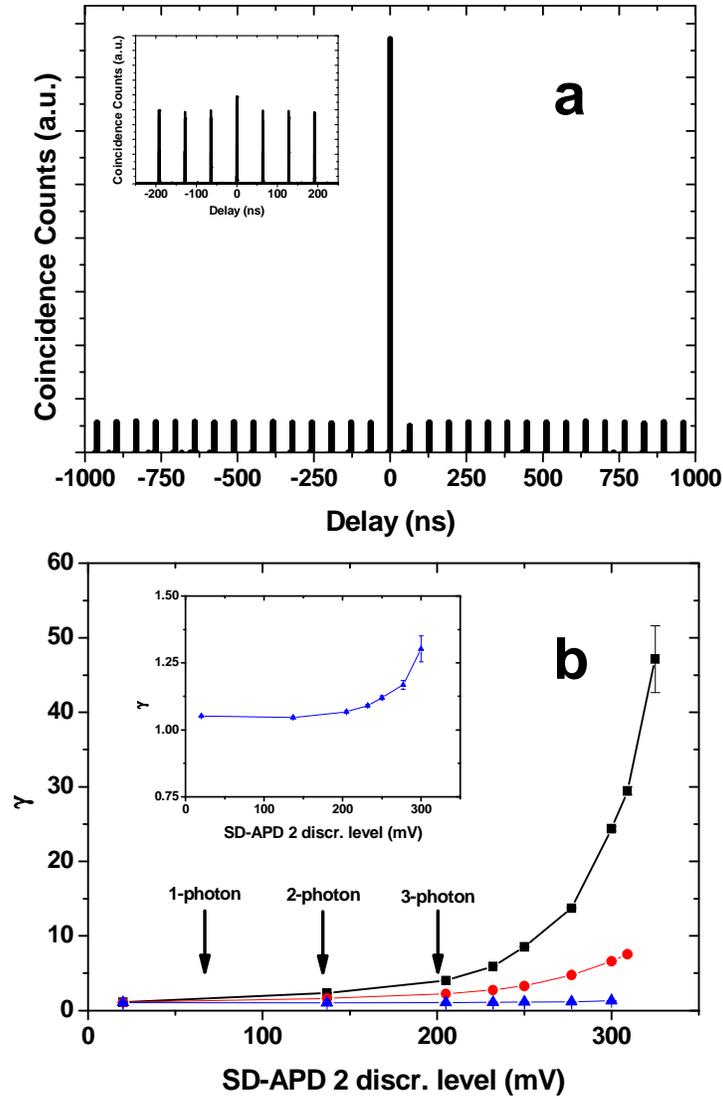}}}}
\caption{\label{fig:fig3.eps}(a) Experimentally measured photon count histogram as a function of detector relative delay for source FML. SD-APD 1 and SD-APD 2 discrimination levels are both set at a high value. Inset: Second order correlation measurement for source FML with $g(2)=1.2$. SD-APD 1 and SD-APD 2 discrimination levels are both set at the lowest possible discrimination level. (b) Photon correlations, $\gamma$ of the three sources shown in Fig. \ref{fig:fig1.eps}(c): FML (squares), LNT (circles) \& LAT (triangles). The arrows correspond to the mean positions of the photon number states assuming a avalanche voltage linear dependence on photon number. Inset: Detail of $\gamma$ for source LAT. Error bars are $\pm$ 1 standard deviation of the plotted (mean) values and are plotted for all data; however only some error bars are visible due to the size of the data points.}
\end{figure}

\begin{figure}[t!]
{\centerline{\scalebox{0.4}{\includegraphics*{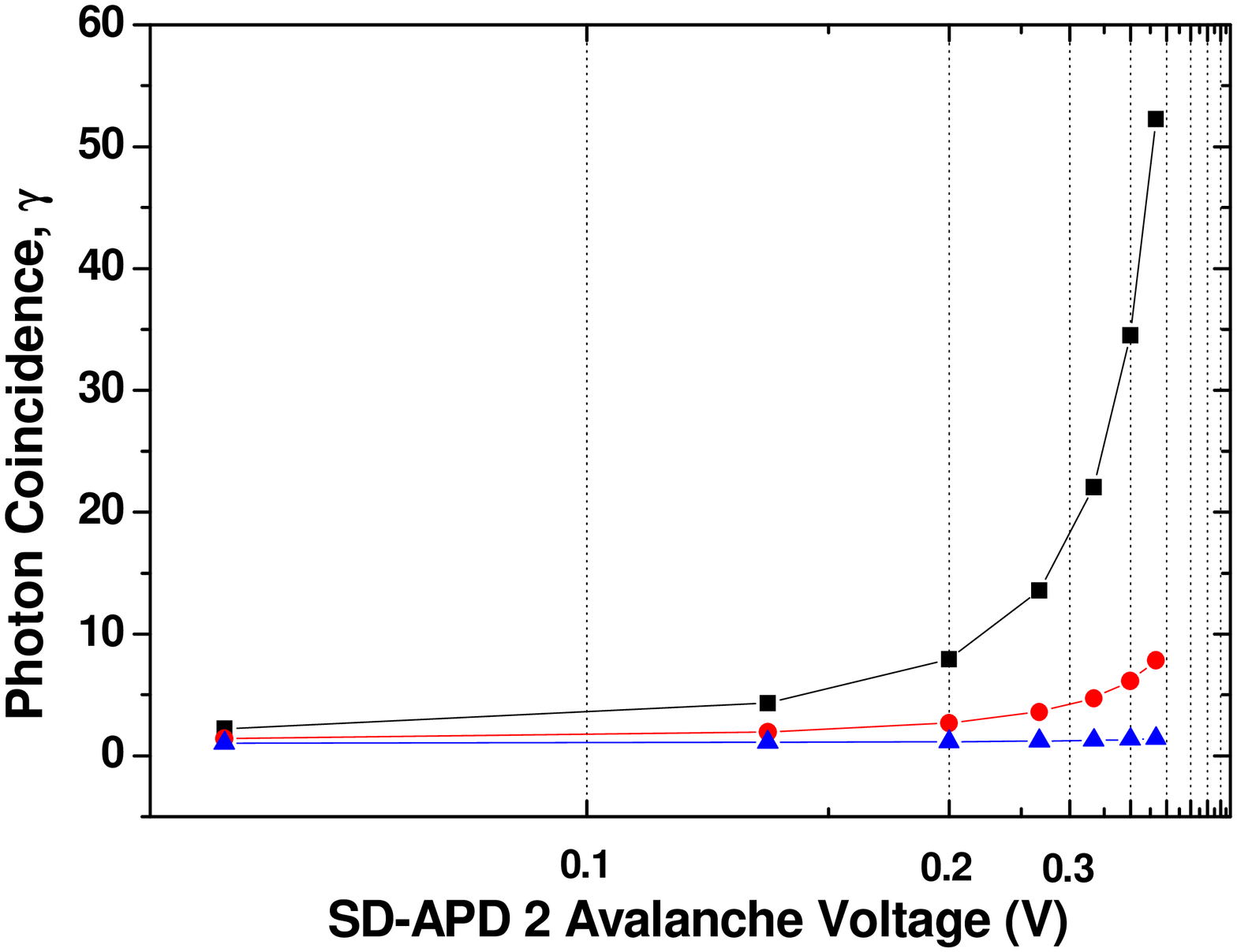}}}}
\caption{\label{fig:fig4.eps} Simulated correlation value as a function of SD-APD 2 discrimination level for SD-APD1 discriminator level set at $n = 7$ photon number state. FML (squares), LNT (circles) \& LAT (triangles).}
\end{figure}

\section{Higher order correlations of the photon field}

Having shown that the detectors are sensitive to differences in the photon number distribution, we now present experimental results that detectors can be used to measure higher order correlations of the photon field.
The setup illustrated in Fig. \ref{fig:fig1.eps}(a) is used.  We stress that the setup will be exactly the same as a conventional one that measures the second order correlation function g(2) if the photon number resolving function is disabled. By setting the discrimination level at ~300mV, a value far greater than single photon generated avalanches, for both detectors SD-APD 1 and SD-APD 2, the measurement gives a correlation result that is different from g(2). Plotted in Fig. \ref{fig:fig3.eps}(a), is a histogram for source FML when both the discrimination levels of SD-APD 1 and SD-APD 2 are at high values. At zero time delay ($\Delta t=0$) a prominent correlation peak is observed due to photon bunching. The correlation here is given by Eq. (4), $\gamma$. The correlation value is measured as the ratio of peak height at t=0 to the average height at $\Delta t\neq0$. $\gamma$ is much higher than the the equivalent $g(2)$ value of $g(2) = 1.2$ , as shown by the histogram in the inset of Fig. \ref{fig:fig3.eps}(a).%$\tilde{g}$ encapsulates the degree of correlation measured.

For different discrimination levels, correlation plots similar to Fig. \ref{fig:fig3.eps}(a) are collected and the correlations, $\gamma$ evaluated. The detector discriminator level of detector SD-APD 1 was kept at $~300$mV. Fig. \ref{fig:fig3.eps}(b) plots the resulting correlations, $\gamma$ as a function of SD-APD 2 discrimination level for the three sources.  All three sources show $\gamma>1$ even at the lowest discrimination level used. %As the discriminator level is tuned to higher levels, $\gamma$ increases in all cases. 
This can be readily understood in terms of elevated photon bunching through higher order correlations. The $n^{th}$ order correlation is the expectation of the joint detections of $n$ photons correlations. For photons that are indistinguishable and statistically dependent, there is a factorial increase of the available permutations of photon amplitudes as $n$ rises \cite{assmann2009,meystre1998}. At higher discrimination levels, higher photon number states feature more markedly than lower photon number states and the correlation is therefore expected to be higher than for lower photon number states. This is in direct contrast to what would be expected for the case of statistically independent photons. In this case the photon statistics are Poissonian and no bunching is possible for any order of correlation.

Photon source LAT shows the least amount of correlation; this is expected for a laser source operating well above threshold with largely Poissonian statistics \cite{milburn2008}. The correlation for LAT is relatively flat over the entire discrimination range of SD-APD 2. $\gamma$ for this source attains $\sim 1.3$ at the very highest D2 discrimination level employed, Fig. \ref{fig:fig3.eps}(b), inset. Operating the same laser near to threshold (LNT) shows a marked increase in correlation which is due to increased intensity fluctuations leading to increased photon bunching. Source FML displays the highest correlation values; attaining $\gamma\sim 47.1\pm4.5$ for these measurements. Intensity fluctuations for this source are by far the highest of all sources under test.

All three sources under study feature different photon number statistics which is incorporated in our model through the mean photon numbers $\mu_{s}$ \& $\mu_{n}$. These mean photon numbers are derived from the total average photon number $\mu$ and $g(2)$. The second order correlation function $g(2)$ was measured separately for each of the sources FML, LNT \& LAT with the values $g(2) = 1.2$, $1.075$ \& $1.001$ respectively (the inset of Fig. \ref{fig:fig3.eps}(a) shows the measured $g(2)$ photon coincidence histogram for source FML). 

As remarked previously, avalanche height self-limiting due to the APD series resistance is likely to distort the linear avalanche voltage dependence on photon number for high photon number. We can incorporate this fact by increasing $n_{max}$ in the model for the detectors. Fig. \ref{fig:fig4.eps} shows the simulated results based on Eq. (\ref{eq:eq2}) for $\gamma$ plotted with up to $n_{max}=7$ photon detection events . Each point corresponds to selecting exactly $n$ avalanche detections based on a linear dependence of the avalanche peaks (as shown by the arrows in Fig. \ref{fig:fig3.eps}(b)). The x-axis is plotted reciprocally to emphasize the avalanche self-limiting. $\gamma$ for all sources shows a similar growth and dependence to that observed experimentally, qualitatively confirming our underlying model.

Finally, we remark that although the agreement between the experimental and theoretical data is good, it is only qualitative. The photon number resolving capability of the detector worsens at higher photon number due to avalanche broadening. Avalanche broadening causes adjacent photon number avalanches to overlap. This fact is not incorporated in the model for simplicity. The next step is to realize a SD-APD with narrow photon number avalanche peaks having minimal overlap; similar to the performance of visible light photon counters (VLPCs), for example \cite{waks2006}. This will permit efficient window discrimination between $n$ and $n\pm1$ photon number avalanche peaks. Then the correspondence between the experimentally measured $\gamma$ and Eq. (4) would be direct.

\section{Conclusion}
In summary we have demonstrated a unique, proof of principle technique for inferring higher order photon correlations based on high speed (GHz) photon number resolving APDs. Using three light sources with differing photon statistics we have shown that higher order correlations of these sources can be detected. Moreover, accessing these higher order correlations illustrates the sensitivity of the method when trying to distinguish between sources compared to a usual $g(2)$ measurement. We believe the technique will be of use and importance in characterizing photon sources involved in the rapidly evolving field of quantum information science.

\section{Acknowledgments}
We thank M. B. Ward for enlightening discussions and a critical reading of the manuscript. This work was partially supported by EU project ``Q-Essence''.

\begin{thebibliography}{99}

\bibitem{glauber1963} R. J. Glauber, ``The quantum theory of optical coherence," Phys. Rev. {\bf{130}}(6), 2529--2539 (1963).

\bibitem{milburn2008} D. F. Walls and G. J. Milburn, ``Coherence properties of the electromagnetic field," in {\it Quantum optics,} (Springer-Verlag, Berlin, 2008), pp. 29--55.

\bibitem{usami2004} K. Usami, Y. Nambu, B. S. Shi, A. Tomita, and K. Nakamura, ``Observation of antinormally ordered Hanbury Brown–-Twiss correlations," \prl {\bf {92}}(11), 113601 (2004).

\bibitem{hanbury1956} R. Hanbury-Brown and R. Q. Twiss, Nature, ``Correlation between photons in two coherent beams of light," Nature {\bf{177}}(4497), 27--29 (1956).

\bibitem{jeltes2007} T. Jeltes, J. M. McNamara, W. Hogervorst, W. Vassen, V. Krachmalnicoff, M. Schellekens, A. Perrin, H. Chang, D. Boiron, A. Aspect, and C. I. Westbrook, ``Comparison of the Hanbury Brown-–Twiss effect for bosons and fermions," Nature {\bf {445}}(7126), 402--405 (2007).

\bibitem{kimble1977} H. J. Kimble, M. Dagenais, and L. Mandel, ``Photon antibunching in resonance fluorescence," \prl {\bf{39}}(11), 691--695 (1977).

\bibitem{agarwal1970} G. S. Agarwal, ``Field--correlation effects in multiphoton absorption processes," \pra {\bf{ 1}}(5), 1445--1459 (1970).

\bibitem{gisin2002_gisin2007} N. Gisin, G. Ribordy, W. Tittel, and H. Zbinden, ``Quantum cryptography," \rmp {\bf 74}(1) 145--195 (2002).

\bibitem{michler2000} P. Michler, A. Kiraz, C. Becher, W. V. Schoenfeld, P. M. Petroff, Lidong Zhang, E. Hu, and A. Imamoglu ``A quantum dot single--photon turnstile device," Science {\bf{290}}(5500), 2282--2285 (2000).

\bibitem{santori2001} C. Santori, M. Pelton, G. Solomon, Y. Dale, and Y. Yamamoto, ``Triggered single photons from a quantum dot," \prl {\bf {86}}(8), 1502--1505 (2001).

\bibitem{yuan2002} Z. Yuan, B. E. Kardynal, R. M. Stevenson, A. J. Shields, C. J. Lobo, K. Cooper, N. S. Beattie, D. A. Ritchie, and M. Pepper, ``Electrically driven single--photon source," Science {\bf{295}}(5552), 102--105 (2002).

\bibitem{adachi2009} Y. Adachi, T. Yamamoto , M. Koashi, and N. Imoto , ``Boosting up quantum key distribution by learning statistics of practical single-photon sources," New J. Phys. {\bf{11}}(11), 113033 (2009).

\bibitem{qu1996} Y. Qu, S. Singh, and C. D. Cantrell, ``Measurements of higher order photon bunching of light beams," \prl {\bf{76}} (8), 1236--1239 (1996).

\bibitem{stevens2009} M. J. Stevens, B. Baek, E. A. Dauler, A. J. Kerman, R. J. Molnar, S. A. Hamilton, K. K. Berggren, R. P. Mirin, and S. W. Nam, ``High--order temporal coherences of chaotic and laser light," Opt. Expr. {\bf{18}}(2), 1430--1437 (2010), \url{http://www.opticsexpress.org/abstract.cfm?URI=OPEX-18-2-1430}.

\bibitem{avenhaus2009} M. Avenhaus, K. Laiho, M. V. Chekhova, and C. Silberhorn, ``Accessing higher order correlations in quantum optical states by time multiplexing," \prl {\bf 104}(6), 063602 (2010).

\bibitem{kalashnikov2011} D. A. Kalashikov, S. H. Tan, M. V. Chekhova, and L. A. Krivitsky, ``Accessing photon bunching with photon number resolving multi-pixel detector,'' Opt. Expr. {\bf{19}}(10), 9352-9363 (2011), \url{http://www.opticsinfobase.org/oe/abstract.cfm?uri=oe-19-10-9352}. %arXiv:1012.2264 [quant-ph] \url{http://arxiv.org/abs/1012.2264} 

\bibitem{hadfield2009} R. H. Hadfield, ``Single-photon detectors for optical quantum information applications,'' Nat. Phot. {\bf {3}} 696--705 (2009)

\bibitem{yuan2007} Z. L. Yuan, B. E. Kardynal, A. W. Sharpe, and A. J. Shields, ``High speed single photon detection in the near infrared," \apl {\bf{91}}(4), 041114 (2007).

\bibitem{kardynal2008} B. E. Kardynal, Z. L. Yuan, A. J. Shields, ``An avalanche--photodiode--based photon--number--resolving detector," Nat. Photon. {\bf{2}}(7), 425--428 (2008).

\bibitem{thomas2010} O. Thomas, Z. L. Yuan, J. F. Dynes and A. J. Shields, ``Efficient photon number detection with silicon avalanche photodiodes,'' \apl {\bf{97}} (3) 031102 (2010)

\bibitem{mandel1995} L. Mandel and E. Wolf, ``Quantum theory of photoelectric light detection," in {\it Optical Coherence and Quantum Optics,} (Cambridge University Press, New York, 1995), pp. 683--740.

\bibitem{dixon2009} A. R. Dixon, J. F. Dynes, Z. L. Yuan, A. W. Sharpe, A. J. Bennett, and A. J. Shields, ``Ultrashort dead time of photon-counting InGaAs avalanche photodiodes," \apl {\bf 94}(23), 231113 (2009).

\bibitem{smith1966} A. W. Smith and J. A. Armstrong, ``Laser photon counting distributions near threshold," \prl {\bf{16}}(25), 1169--1172 (1966);

\bibitem{lachs1965} G. Lachs, ``Theoretical aspects of mixtures of thermal and coherent radiation," Phys. Rev. {\bf {138}}(4B), B1012--B1016 (1965).

\bibitem{g2ref} The value of $g(2) \sim 1.2$ is corroborated experimentally by an independent measurement of $g(2)$.

\bibitem{assmann2009} M. A\ss mann, F. Veit, M. Bayer, M. van der Poel, and J. M. Hvam, ``Higher-order photon bunching in a semiconductor microcavity," Science {\bf{325}}(5938), 297--300 (2009)

\bibitem{meystre1998} P. Meystre and M. Sargent, ``Field quantization," in {\it Elements of Quantum Optics,} (Springer--Verlag, Berlin, 1998), pp. 263--285.

\bibitem{waks2006} E. Waks. E. Diamanti and Y. Yamamoto ``Generation of photon number states,'' New J. Phys. {\bf{8}} (1), 4 (2006)

\end{thebibliography}
\end{document}